\documentclass[aps,pre,reprint,longbibliography,nofootinbib,floatfix]{revtex4-2}
\usepackage{amsmath,amssymb}
\usepackage{graphicx,booktabs,capt-of}
\usepackage{microtype}
\usepackage[hidelinks,pdfauthor={Reon Hayashi},
  pdftitle={Memory-induced optimal switching in a dynamic Ising-Kuramoto model}]{hyperref}
\newcommand{\dd}{\mathrm{d}}
\newcommand{\kc}{K_{\mathrm c}}
\newcommand{\av}[1]{\left\langle #1\right\rangle_0}
\begin{document}
\raggedbottom
\brokenpenalty=10000
\title{Memory-induced optimal switching in a dynamic Ising--Kuramoto model}
\author{Reon Hayashi}
\affiliation{Hiroshima Prefectural Fukuyama Seishikan Senior High School, 6-11-1 Kinosho-cho, Fukuyama, Hiroshima 720-0082, Japan}
\begin{abstract}
We study how internal spin switching competes with phase decorrelation
in a globally coupled dynamic Ising--Kuramoto model. Linear response
about the nonmagnetic incoherent state gives
\(K_c=[(\mu+\beta_s\gamma)\chi_\gamma]^{-1}\), where
\(\chi_\gamma\) is a switching-weighted Laplace transform of the
free phase correlation. Under explicit assumptions on that correlation,
we establish sufficient conditions for the mean-field linear instability
threshold to attain a global minimum at a finite positive switching rate,
strictly below both endpoint thresholds. The criterion separates low-rate
integrated phase memory from high-rate short-time correlation structure.
Finite-correlation-time Ornstein--Uhlenbeck forcing satisfies the
high-rate condition; applying the sufficient criterion still requires
checking the low-rate moment condition. The white-noise threshold is
monotone or exactly flat. Finite-system
simulations exhibit appreciable spin-weighted coherence, weak ordinary
coherence and a strong antiphase sector bias. On a magnetized background,
amplitude ratios are quantitatively consistent with the predicted
linear-mode composition locally near onset. Direct finite-size tests
of the nonmagnetic threshold minimum remain unresolved under the
prespecified support rule, despite the predicted conditional contrast
signs at two tested OU times.
\end{abstract}

\maketitle
\section{Introduction}\label{sec:introduction}

Phase-oscillator models provide a standard framework for collective
synchronization~\cite{kuramoto1984}. When the interacting elements also
carry discrete internal states, their response depends on both phase
evolution and internal-state relaxation. Coupled XY--Ising models combine
discrete and continuous ordering~\cite{granato1991}, and oscillator
phases can coevolve with binary interaction strategies~\cite{antonioni2017}.
These settings motivate the question of how an internal relaxation rate
competes with phase decorrelation in the onset of collective coherence.

Several ingredients of this problem have established precedents.
In models with quenched binary interactions, weighted order parameters
already describe coherence that ordinary phase averaging can
miss~\cite{kloumann2014}. Linear-stability analyses have established
how temporally correlated frequency fluctuations modify synchronization
thresholds~\cite{rougemont2006,tonjes2010}. Oscillator susceptibility
has been studied under external forcing~\cite{daido2015}, and
Laplace-domain response formulations have been developed about
nonsynchronized states~\cite{terada2020}. Favorable finite switching
windows are also known in stochastic oscillator networks~\cite{jeter2015},
alongside prescribed on--off coupling and temporal-network synchronization
advantages~\cite{chen2009,zhang2021}. The contribution here is the
combined endogenous spin and spin-correlated phase response, together
with a sufficient criterion for a finite positive global minimum of
its linear instability threshold.

The question addressed here is more specific: under what properties of
the free phase correlation does the linear instability threshold of a
system with endogenous spin reversals necessarily attain a finite-rate
global minimum? A correlation-function formulation separates the
response generated by spin switching from the phase decorrelation that
limits its temporal accumulation. It also distinguishes conditions on
the entire correlation function from those on its short-time behavior.

We consider a globally coupled dynamic Ising--Kuramoto model with
phase-dependent Glauber spin reversals and OU phase forcing. Its
spin-weighted coherence is
\(Z_m=N^{-1}\sum_j\sigma_j e^{i\theta_j}\).
For the nonmagnetic incoherent background, linear response gives
\[
 K_c(\gamma)=[(\mu+\beta_s\gamma)\chi_\gamma]^{-1},
 \qquad
 \chi_\gamma=\frac12\int_0^\infty e^{-\gamma t}G(t)\,\dd t.
\]
Under the stated assumptions on \(G\), two strict endpoint conditions
ensure a global minimum at a finite positive switching rate, strictly
below both endpoint thresholds. This sufficient criterion combines
low-rate integrated memory with high-rate short-time correlation
structure. Finite-correlation-time OU forcing satisfies the high-rate
condition, while the low-rate condition must still be checked to apply
the sufficient criterion.
The white-noise threshold is instead monotone or
flat. We also derive the ordinary-coherence projection of the composite
linear mode on a magnetized background.

Simulations examine finite-size consequences of these analytical results.
The sampled nonmagnetic systems exhibit appreciable spin-weighted
coherence, weak ordinary coherence and an antiphase sector bias. On a
magnetized background, amplitude ratios are quantitatively consistent
with the linear-mode prediction locally near onset. Direct finite-size
tests of nonmagnetic threshold nonmonotonicity remain unresolved under
the prespecified support rule, although two tested OU times yield the
predicted conditional contrast signs. The analytical criterion and these
finite-system observations therefore have distinct evidential roles.

\clearpage

\section{Model and conventions}\label{sec:model}

Each of \(N\) globally coupled units carries a phase \(\theta_i\) modulo
\(2\pi\) and an Ising spin \(\sigma_i=\pm1\). We use the magnetization \(m\),
the ordinary phase coherence \(Z\), and the spin-weighted coherence \(Z_m\),
\begin{align}
 m&=\frac1N\sum_j\sigma_j,\qquad Z=\frac1N\sum_j e^{i\theta_j},\nonumber\\
 Z_m&=\frac1N\sum_j\sigma_j e^{i\theta_j},\qquad
 r=|Z|,\quad r_m=|Z_m|. \label{eq:ops}
\end{align}
With equal populations and coherence amplitudes, antiphase sectors
cancel in \(Z\) and add in \(Z_m\). The interaction energy is
\begin{equation}
 \mathcal H=-\frac{1}{2N}\sum_{i,j}\sigma_i\sigma_j
               [J+K\cos(\theta_i-\theta_j)] . \label{eq:energy}
\end{equation}
Here \(J\) is the ferromagnetic coupling and \(K>0\) the spin-dependent
phase coupling. The ordinary phase-only coupling is zero (\(\lambda=0\)).
The factor \(1/2\) counts each pair once, and the diagonal terms are constant.

The phase drift is \(-\mu\,\partial_{\theta_i}\mathcal H\), with mobility
\(\mu>0\). Each phase also experiences independent stationary
Ornstein--Uhlenbeck (OU) forcing~\cite{uhlenbeck1930},
\begin{align}
 \dot\theta_i&=\mu K\sigma_i\,\operatorname{Im}(Z_m e^{-i\theta_i})+\eta_i,
       \label{eq:phase}\\
 \tau\,\dd\eta_i&=-\eta_i\,\dd t+\sqrt{2D}\,\dd W_i. \label{eq:ou}
\end{align}
The \(W_i\) are independent standard Wiener processes, \(\tau>0\) is the
correlation time, and \(D>0\) is the one-sided integrated intensity,
\begin{equation}
 C_\eta(u)=\frac{D}{\tau}e^{-|u|/\tau},\qquad
 \int_0^\infty C_\eta(u)\,\dd u=D. \label{eq:noise}
\end{equation}
Varying \(\tau\) at fixed \(D\) also changes the instantaneous variance
\(D/\tau\).

The Glauber flip rate~\cite{glauber1963} at inverse spin temperature
\(\beta_s>0\) is
\begin{equation}
 w_i(\sigma_i\to-\sigma_i)
 =\frac{\gamma}{2}[1-\sigma_i\tanh(\beta_s h_i)]. \label{eq:flip}
\end{equation}
At finite \(N\) a flip costs \(2\sigma_i h_i\) in interaction energy.
Removing the diagonal contributions in Eq.~\eqref{eq:energy} gives the
field used in the simulations,
\begin{equation}
 h_i=Jm+K\operatorname{Re}(Z_m e^{-i\theta_i})
                  -\frac{J+K}{N}\sigma_i. \label{eq:field}
\end{equation}
The mean-field theory takes \(N\to\infty\), leaving the first two terms.
At zero field the one-way flip rate is \(\gamma/2\), and the single-spin
correlation decays at rate \(\gamma\).

\section{Linear instability and finite-rate threshold minima}\label{sec:theory}
\subsection{Composite instability of the nonmagnetic state}

In the incoherent state with \(m=Z=Z_m=0\), the uniform free phase
\(\theta_0\) is independent of the unbiased spin. We take \(\gamma>0\) and \(\beta_s J<1\), so uniform
magnetic perturbations decay. The free correlation and susceptibility are
\begin{align}
 G(t)&=\av{e^{i[\theta_0(t)-\theta_0(0)]}},\nonumber\\
 \chi_a&=\frac12\int_0^\infty e^{-at}G(t)\,\dd t. \label{eq:chi}
\end{align}
The average is over the stationary free process. Under the integrability
assumptions stated below, \(\chi_a\) is holomorphic for \(\Re a>0\)
and continuous for \(\Re a\geq0\); the finite first moment also gives
its right-sided boundary derivative.
In the nonmagnetic characteristic equation, \(a=\gamma+s\) with
\(\gamma>0\) and \(\Re s\geq0\), so no analytic continuation across
the imaginary axis is used. For a perturbation
\(Z_m(t)\propto e^{st}\) with complex growth exponent \(s\), the
phase drift contributes \(\mu K\chi_{\gamma+s}Z_m\).
The field-dependent bias in the spin flips contributes
\(\beta_s\gamma K\chi_{\gamma+s}Z_m\). The same damping in these two
terms comes from the free spin correlation in the first term and the
relaxation of the driven spin mean in the second (Appendix~\ref{app:nonmag}).
Self-consistency gives
\begin{equation}
 1=K(\mu+\beta_s\gamma)\chi_{\gamma+s}. \label{eq:char}
\end{equation}
For a real positive \(G\) the first marginal mode in this sector is
stationary. The linear threshold is therefore
\begin{equation}
 \kc(\gamma)=\frac{1}{(\mu+\beta_s\gamma)\chi_\gamma}. \label{eq:kc}
\end{equation}
Increasing \(\gamma\) raises the Glauber prefactor \(\beta_s\gamma\)
while shortening the time over which a phase perturbation contributes
to the susceptibility.

\subsection{A sufficient condition for a finite minimum}

Hold the free phase law fixed as \(\gamma\) varies. Define
\begin{align}
 M_n(\gamma)&=\int_0^\infty t^n e^{-\gamma t}G(t)\,\dd t,\nonumber\\
 F(\gamma)&=(\mu+\beta_s\gamma)M_0(\gamma)
             =\frac{2}{\kc(\gamma)}. \label{eq:F}
\end{align}
The value at \(\gamma=0\) denotes the continuous right-limit extension
of the threshold. At zero switching the spins are frozen.

\textit{Proposition.} Suppose \(G\) is real, positive and measurable,
\(M_0(0), M_1(0)<\infty\), and \(G(0)=1\) with
\(G(t)=1+g t+O(t^2)\) as \(t\downarrow0\), where \(g=G'(0^+)\).
If
\begin{equation}
 \beta_s M_0(0)>\mu M_1(0),\qquad
 \mu+\beta_s g>0, \label{eq:margins}
\end{equation}
then \(\kc\) attains its global minimum at a finite positive switching
rate. The minimum lies strictly below both endpoint values.

The first inequality makes \(\kc\) decrease as switching begins. It
depends on the entire free correlation. The second makes \(\kc\)
increase at sufficiently large rates, where the susceptibility samples
only short times. Appendix~\ref{app:minimum} proves attainment and treats
the marginal signs. The criterion is sufficient and leaves the number of
minimizers open. At any interior stationary rate \(\gamma_\ast\),
\begin{equation}
 \frac{M_1(\gamma_\ast)}{M_0(\gamma_\ast)}
       =\frac{\beta_s}{\mu+\beta_s\gamma_\ast}. \label{eq:balance}
\end{equation}
The relevant time on the left is a mean lag weighted by both phase
correlation and spin relaxation.

\subsection{OU forcing and the white-noise limit}

Let \(\Delta\theta_0(t)=\int_0^t\eta(u)\,\dd u\) denote the
\emph{unwrapped} free phase increment. Integrating the stationary
covariance in Eq.~\eqref{eq:noise} gives
\begin{align}
 \operatorname{Var}\Delta\theta_0(t)
 &=2\int_0^t(t-u)C_\eta(u)\,\dd u\nonumber\\
 &=2D[t-\tau(1-e^{-t/\tau})]. \label{eq:var}
\end{align}
This increment is centered Gaussian. Its first harmonic depends only
on the increment modulo \(2\pi\),
\begin{equation}
 G(t)=e^{-\operatorname{Var}\Delta\theta_0(t)/2}
 =\exp\{-D[t-\tau(1-e^{-t/\tau})]\}. \label{eq:G}
\end{equation}
For each finite \(\tau>0\), \(G(t)=1-Dt^2/(2\tau)+O(t^3)\).
Thus \(g=0\), and \(\mu>0\) ensures the high-rate inequality.
The threshold approaches \(2/\beta_s\) from below at large \(\gamma\).
A finite-rate minimum follows when the low-rate moment inequality in
Eq.~\eqref{eq:margins} also holds.

As \(\tau\downarrow0\) at fixed \(D\), \(C_\eta\to2D\delta\) on the
full time axis and \(G(t)\to e^{-Dt}\) for \(t\geq0\). The white-noise
susceptibility is \(\chi_a=[2(D+a)]^{-1}\), giving
\begin{align}
 \kc^{\rm white}(\gamma)&=\frac{2(D+\gamma)}{\mu+\beta_s\gamma},\nonumber\\
 \frac{\dd \kc^{\rm white}}{\dd\gamma}
       &=\frac{2(\mu-\beta_s D)}{(\mu+\beta_s\gamma)^2}. \label{eq:white}
\end{align}
The threshold increases for \(\mu>\beta_s D\) and decreases for
\(\mu<\beta_s D\). At equality every rate gives the same threshold
\(2/\beta_s\).

\subsection{Magnetized incoherent background}

For \(\beta_sJ>1\) let \(m_0\) lie on either stable magnetized branch,
\begin{equation}
 m_0=\tanh(\beta_s Jm_0),\qquad
 \beta_s J(1-m_0^2)<1. \label{eq:m0}
\end{equation}
The background is uniform in phase, with \(Z=Z_m=0\). The free spin
correlation now has a constant part \(m_0^2\), which carries phase
memory even after the connected spin correlation decays.
Appendix~\ref{app:magnet} gives the stationary composite threshold
\begin{equation}
 \kc^{(m_0)}
 =\big[\mu m_0^2\chi_0+
 (1-m_0^2)(\mu+\beta_s\gamma)\chi_\gamma\big]^{-1}.
 \label{eq:magkc}
\end{equation}
The marginal eigenvector also carries ordinary phase coherence. Its
amplitude ratio at linear onset is
\begin{equation}
 \left(\frac{r}{r_m}\right)_{\rm linear\ onset}
     =\mu\kc^{(m_0)}|m_0|\chi_0. \label{eq:magratio}
\end{equation}
Reversing the magnetization reverses the signed projection \(Z/Z_m\)
and preserves its magnitude.

\section{Numerical design and diagnostics}\label{sec:numerics}

We sampled the nonmagnetic and magnetized regimes in separate
campaigns (Table~\ref{tab:design}). Both used \(\lambda=0\),
\(\beta_s=2\), and \(D=\mu=1\). At each site we drew the initial phase
uniformly on the circle, the spin with equal probabilities, and the
OU forcing from its stationary Gaussian law of variance \(D/\tau\).
These initial draws were independent. After 200 time units we recorded
observables every 0.05 for 500 time units. We first averaged within
each realization, then over realizations.
Here ``prespecified'' means fixed before the clean production campaign;
no external registration is claimed.

\begin{center}\begin{minipage}{\columnwidth}
\captionof{table}{\label{tab:design}Production designs. For the \(J=0.4\)
crossings, \(N=1000,2000\) are the primary sizes. The \(N=500\)
sensitivity check covers the three prespecified contrast rates at
each \(\tau\). The separate composite-mode panel uses all three sizes.}
\centering
\begin{tabular}{@{}lcc@{}}
\toprule
 & Nonmagnetic & Magnetized \\
\midrule
\(J\) & \(0.4\) & \(0.8\) \\
\(\tau\) & \(0.75,1,1.5\) & \(1\) \\
\(\gamma\) & 7 rates per \(\tau\) & \(1\) \\
\(N\) & \(500,1000,2000\) & \(500,1000,2000\) \\
\(\Delta t\) & \(0.0025\) & \(0.005\) \\
Realizations per point & 12 & 20 \\
Bootstrap draws & 5000 & 10\,000 \\
\bottomrule
\end{tabular}
\end{minipage}\end{center}

At each step we sampled the OU endpoint and the integrated forcing
jointly from their exact Gaussian law conditional on the current
forcing [Eq.~\eqref{eq:ou}]. A Heun step for Eq.~\eqref{eq:phase}
used the same integrated forcing and the old spins in both stages.
We wrapped the corrected phases modulo \(2\pi\), evaluated the rates
with those phases, the old spins and the self-excluded field of
Eq.~\eqref{eq:field}, and flipped spins simultaneously with
probabilities \(1-e^{-w_i\Delta t}\). This spin update approximates
the continuous-time Glauber process of Eq.~\eqref{eq:flip}, whose
one-way zero-field rate is \(\gamma/2\) and whose zero-field
correlation decay rate is \(\gamma\).

\subsection{Finite-size crossing diagnostic}

For \(J=0.4\), each replicate reused its random stream across \(K\),
preserving the pairing of its entire coupling curve. Separate streams
were assigned by campaign stage, \(\tau,\gamma,N,\Delta t\), and
replicate. A threshold-level screen compared \(\Delta t=0.005\) and
\(0.0025\) in one neighborhood at each \(\tau\). All three screens
remained unresolved after expansion to 12 replicates. The prespecified
fallback selected the finer step, leaving its residual discretization
error undetermined.

We used \(r_m\sim N^{-1/4}\) as a working mean-field assumption to
define a pseudocritical diagnostic \(K_\times\). At each size we fitted
the values of \(N^{1/4}\overline{r_m}\) versus \(K\) by equal-weight
nondecreasing isotonic regression using the pool-adjacent-violators
algorithm. Here the bar denotes the time and replicate average.
We linearly interpolated the difference of the two fitted curves.
A single zero gave \(K_\times\), and a single zero plateau gave its
midpoint. Multiple disjoint zero components were ambiguous, and no
zero gave a missing estimate. The coupling grids were centered on
Eq.~\eqref{eq:kc} and extended by a fixed bracketing rule in steps
of \(0.005\), up to \(0.030\) on either side. The exponent and primary
size pair were fixed before production.
Because the coupling grids and local-fit windows were centered on
analytical predictions, proximity of retained crossings or accepted
vertices to theory is not by itself an independent accuracy test.

We resampled complete replicate curves across \(K\), independently
for each size and switching rate. The crossing intervals use the
2.5th and 97.5th percentiles of valid draws. A supported point
required a single zero component, an exchange of ordering, both
grid boundaries more than one step from that component, and a valid
fraction of at least \(0.9\). At each \(\tau\) we tested whether the
middle of three prespecified rates had a lower crossing than both
outer rates. Each difference used draws valid at both rates in that
comparison. The decision required support at all three rates and a
positive lower percentile endpoint for each difference. The Numerical
Supplement gives the rate grids, failure counts, and sensitivity
checks for size, exponent, and interpolation.
Numerical coupling estimates and interval endpoints are displayed to
five decimal places and bootstrap fractions to four; grid settings and
analytical references retain their stated precision.
Interpolation permits sub-grid values without implying corresponding
physical accuracy; full-precision values are retained in the numerical
records.

\subsection{Magnetized mode ratio and sampling uncertainty}

For \(J=0.8\) we sampled five couplings from \(1.095\) to \(1.115\)
in steps of \(0.005\). A six-seed pilot at \(N=500,\ K=1.105\)
preceded the choice of \(\Delta t=0.005\). For production we reused
seed labels \(0,\ldots,19\) across both \(K\) and \(N\). To compare
with Eq.~\eqref{eq:magratio}, we formed
\(R=\overline r/\overline{r_m}\). Within each size we resampled
seed labels jointly for the numerator and denominator and across
all five couplings. Error bars are pointwise percentile intervals.

In both campaigns the sampling unit for uncertainty was a realization.
Split-half, block and autocorrelation diagnostics assessed temporal
variability within runs. All trajectories were retained. The
bootstrap intervals quantify finite-seed variability at fixed
settings. Effects of system size, time discretization, and finite
duration require separate assessment. Convergence with burn or
sampling duration and independence from the initial-state
distribution remain unestablished.

\section{Finite-size comparisons}\label{sec:comparisons}
\subsection{Analytical threshold and finite-size crossings}

\begin{figure*}[t!]
\centering
\includegraphics[width=0.73\textwidth]{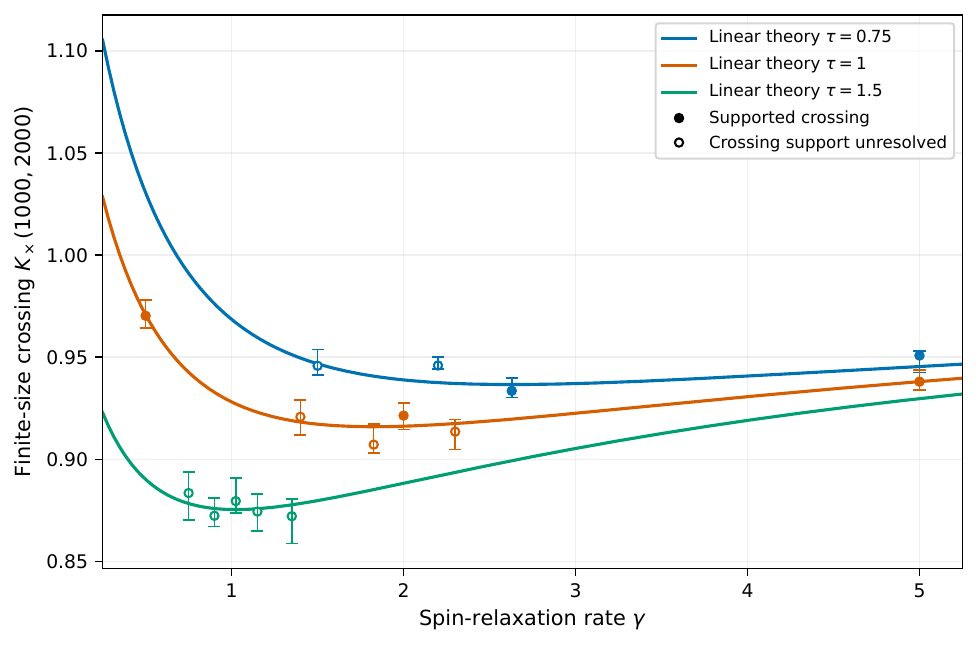}
\caption{\label{fig:thresholds}Stationary linear threshold and finite-size
crossings at \(J=0.4,\ D=\mu=1,\ \beta_s=2\).
The colored curves are Eq.~\eqref{eq:kc} at the indicated OU times.
Symbols are crossings of the two size-specific, isotonic fits to
\(N^{1/4}\overline{r_m}\) at \(N=1000,2000\), using the simulation
means and linear interpolation. Filled symbols meet the prespecified
support criterion. Open symbols have a valid point crossing with
insufficient support. Bars give pointwise 95\% bootstrap intervals
conditional on a valid crossing. Six ambiguous point estimates are
omitted. Only the analytical curves connect rates. The points are
pseudocritical diagnostics under a working scaling assumption, with
an undetermined extrapolation to the thermodynamic limit. The
conditional contrasts at \(\tau=0.75,1.0\) have the sign predicted for
a finite-rate dip, but the direct test remains unresolved for all three
\(\tau\) under the complete support rule (Table~\ref{tab:verdict}).}
\end{figure*}

The conditional contrasts have the sign predicted for a finite-rate dip
at \(\tau=0.75,1.0\), but the prespecified direct finite-size tests
remain unresolved at all three OU times (Table~\ref{tab:verdict}).
Figure~\ref{fig:thresholds} compares the stationary linear threshold
of Eq.~\eqref{eq:kc} with the crossings \(K_\times(1000,2000)\).
Of 21 sampled \((\tau,\gamma)\) conditions, 15 gave a single crossing
component in the point estimates. Six gave disjoint components, and
only five of the 21 met the full support criterion in
Sec.~\ref{sec:numerics}.

\begin{table*}[t!]
\caption{\label{tab:verdict}Direct test of a lower pseudocritical
coupling at the middle of three prespecified switching rates,
\(J=0.4\). Each entry in the two contrast columns is a pointwise
95\% bootstrap interval conditional on usable crossings at both rates
in that contrast. The decision also requires supported point crossings
at all three rates, including a valid-draw fraction \(f\geq0.9\) at
each.}
\centering
\small
\begin{tabular}{@{}lcccl@{}}
\toprule
\(\tau\) & \(K_{\times,\rm low}-K_{\times,\rm mid}\) &
 \(K_{\times,\rm high}-K_{\times,\rm mid}\) & Crossing support limitation & Decision\\
\midrule
0.75 & \([0.00535,\,0.02126]\) & \([0.00739,\,0.02125]\) &
 low \(f=0.5656\) & Unresolved\\
1.0  & \([0.05086,\,0.07232]\) & \([0.01971,\,0.03818]\) &
 mid \(f=0.7124\) & Unresolved\\
1.5  & \([-0.00054,\,0.03056]\) & \([-0.00095,\,0.02044]\) &
 \shortstack{low/high ambiguous,\\mid \(f=0.8618\)} & Unresolved\\
\bottomrule
\end{tabular}
\end{table*}

At \(\tau=0.75\) and \(1.0\), both pointwise conditional intervals for
the outer-minus-middle contrasts lie entirely above zero.
The prespecified crossing-support criterion nevertheless fails at the
low rate for \(0.75\) and at the middle rate for \(1.0\).
At \(1.5\) both intervals include zero and the low- and
high-rate point crossings are ambiguous. Changing the size pair also
changes the diagnostic. At the middle rates for \(\tau=1.0\) and
\(1.5\), replacing \((1000,2000)\) by \((500,1000)\) shifts
\(K_\times\) by \(+0.01268\) and \(-0.00858\), respectively.

Under Eq.~\eqref{eq:margins}, the mean-field linear threshold attains a
global minimum at a finite positive switching rate, strictly below both
endpoint thresholds (Sec.~\ref{sec:theory}).
The finite-size test assesses a dip at three specified rates.
The Numerical Supplement describes
the vertices of accepted local fits without treating them as an
independent detection of an optimum. Their fitted ordinate \(K_{\min}\)
is an absolute pseudocritical coupling. Neither that ordinate nor
the fixed-rate contrasts in Table~\ref{tab:verdict} measure the
optimized reduction relative to a baseline.

\subsection{Composite-mode structure in finite systems}

At \(J=0.4,\ \tau=\gamma=1\), the sampled finite systems exhibit a
clear composite-mode structure: spin-weighted coherence substantially
exceeds ordinary coherence over the sampled coupling range.
For example, at \(K=0.930,\ N=2000\) the means were
\(\overline r=0.0206\) and \(\overline{r_m}=0.1620\).
At the same point the mean cosine of the phase difference between
spin sectors was \(-0.949\).

\begin{figure*}[t!]
\centering
\includegraphics[width=0.87\textwidth]{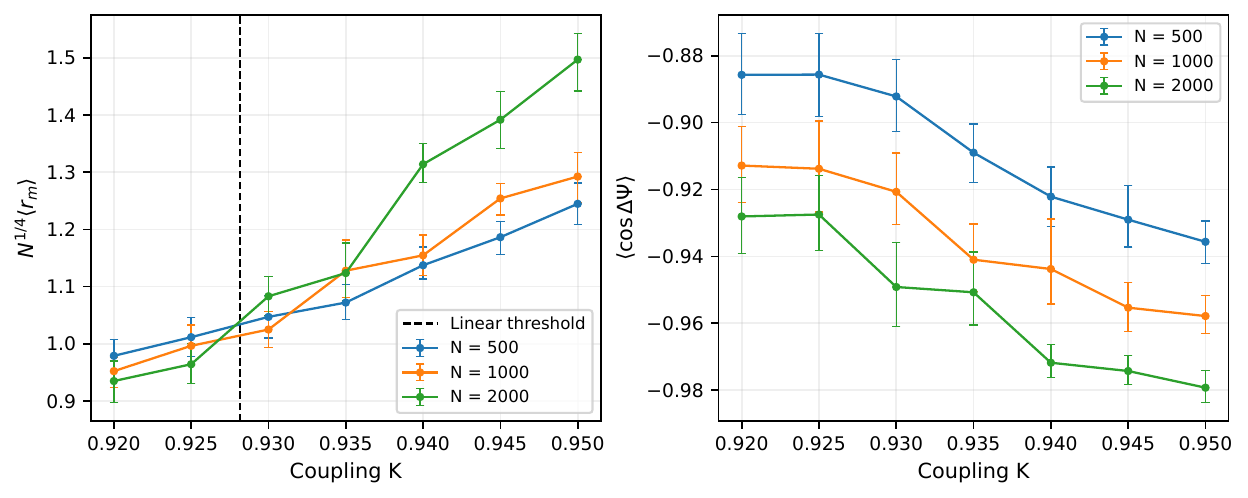}
\caption{\label{fig:sectors}Composite coherence and relative sector
phase in finite systems at \(J=0.4,\ \tau=\gamma=D=\mu=1,\ \beta_s=2\).
Left, \(N^{1/4}\overline{r_m}\) at three sizes. The vertical dashed
line is the analytical threshold from Eq.~\eqref{eq:kc}.
Right, \(\overline{\cos\Delta\Psi}\), where \(\Delta\Psi\) is the
difference between the arguments of the two spin sectors' mean phase
vectors. Each symbol gives an equally weighted mean of 12 simulation
time averages. Bars are pointwise 95\% percentile intervals from
5000 realization-level bootstrap draws at fixed numerical settings;
both observables and all couplings are resampled jointly within each
size (Numerical Supplement, Sec.~S1).
Lines joining symbols guide the eye. The exponent
\(1/4\) sets the working scale. The scaled point means exchange ordering
more than once. The finite systems show appreciable spin-weighted
coherence and a strong antiphase bias of the two spin sectors; these
fixed-rate observations do not locate a unique crossing.}
\end{figure*}

\noindent\begin{minipage}{\columnwidth}
Its strongly negative values across the panel
indicate that the sector phases favor antiphase, consistent with their
contributions adding in \(Z_m\) and largely cancelling in \(Z\)
(Fig.~\ref{fig:sectors}). The scaled point means at \(N=1000\) and \(2000\)
exchange ordering repeatedly, including between \(K=0.930\) and
\(0.935\) and again between \(0.935\) and \(0.940\).
The pointwise intervals at \(K=0.930,0.935\) overlap between these
sizes; these small reversals are not treated as statistically resolved
crossings.
\end{minipage}

\begin{figure*}[tp!]
\centering
\includegraphics[width=0.44\textwidth]{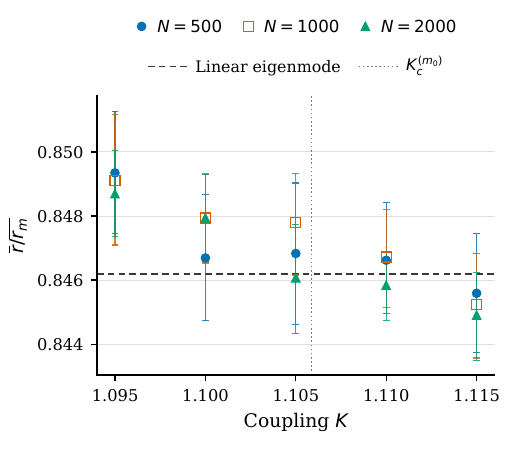}
\caption{\label{fig:magnetic}Ordinary to spin-weighted coherence on
the magnetized background at \(J=0.8,\ \tau=\gamma=D=\mu=1,\ \beta_s=2\).
Each simulation symbol is the ratio \(\overline r/\overline{r_m}\)
of two means over time and 20 realizations. Bars are pointwise 95\%
intervals from joint seed resampling of numerator and denominator.
They quantify seed variability at the recorded numerical settings and
do not include systematic discretization, finite-size, initialization,
or finite-time uncertainty.
The vertical dotted line marks the analytical onset \(1.1058624\).
The horizontal dashed line gives the linear eigenvector ratio
\(0.846193\) at that onset. The simulation symbols are unconnected.
All three size intervals include the reference at \(K=1.105,1.110\)
and lie above it at \(K=1.095\). This is a local quantitative finite-size
check of the predicted mode composition near the analytical onset at
fixed switching rate.}
\end{figure*}

\noindent\begin{minipage}{\columnwidth}
\subsection{Magnetized-background mode ratio}

For \(J=0.8\) the predicted onset is \(\kc^{(m_0)}=1.1058624\),
with a linear-mode amplitude ratio of \(0.846193\).
At both \(K=1.105\) and \(1.110\) the predicted ratio lies within the
pointwise bootstrap interval for each of the three sizes
(Fig.~\ref{fig:magnetic}). At \(K=1.095\) all three intervals lie
above the reference. This local near-onset agreement provides a
quantitative finite-size check of the predicted magnetized linear-mode
composition. The observable compares amplitudes averaged over a finite
sampling time with the ratio of the marginal eigenvector components.
\end{minipage}

\clearpage
\section{Discussion}\label{sec:discussion}

The finite-rate minimum is a property of the combined spin and phase
response. In Eq.~\eqref{eq:kc}, increasing \(\gamma\) strengthens the
spin-response prefactor \(\mu+\beta_s\gamma\), but the factor
\(e^{-\gamma t}\) in \(\chi_\gamma\) restricts the time over which
phase correlations contribute. The two inequalities in
Eq.~\eqref{eq:margins} identify when this competition necessarily
produces a global minimum of the mean-field linear threshold at a finite
positive rate, below both endpoint thresholds. At low rates,
\(\beta_sM_0(0)>\mu M_1(0)\) means that the response gain outweighs
the loss from shortening the correlation window. This condition depends
on integrals over the entire free phase correlation. At high rates,
only its short-time structure enters the leading derivative:
\(\mu+\beta_sG'(0^+)>0\) makes the threshold rise toward its
fast-switching limit. The existence criterion therefore involves two
distinct aspects of the phase dynamics, rather than a single comparison
of characteristic times.
The scalar criterion applies when a system's linear response reduces
to the stated positive-correlation kernel; matching \(G(t)\) alone
is not sufficient.
The derivation uses additive free phase forcing, independent reference
spin and phase processes, an imposed field that leaves the free forcing
law unchanged, and a free phase law held fixed as \(\gamma\) varies.

Finite-correlation-time OU forcing makes this distinction explicit.
The free phase correlation initially decays quadratically, so
\(G'(0^+)=0\) and the high-rate inequality holds for \(\mu>0\).
The low-rate moment inequality must still hold to apply this sufficient
criterion; colored forcing alone
does not guarantee an optimum. For white forcing, the positive-time
correlation is \(G(t)=e^{-Dt}\), whose even extension has a linear
cusp at the origin. Its derivative instead contributes \(-\beta_sD\)
to the high-rate margin. The resulting threshold in
Eq.~\eqref{eq:white} is monotone unless \(\mu=\beta_sD\), when it
is exactly flat. The OU-to-white comparison thus identifies a specific
short-time effect in the susceptibility. It should not be interpreted
as varying memory alone: at fixed integrated intensity \(D\), varying
\(\tau\) also changes the OU variance \(D/\tau\).

The finite-system mode structure provides a direct physical counterpart
to the composite response. Weighting each phase vector by its spin in
\(Z_m=N^{-1}\sum_j\sigma_j e^{i\theta_j}\) makes antiphase
contributions from opposite spin sectors add, whereas they tend to
cancel in the ordinary coherence \(Z\). The appreciable \(r_m\),
much smaller \(r\), and strongly negative sector-phase cosines in the
sampled nonmagnetic systems are consistent with this structure
(Fig.~\ref{fig:sectors}). The joint observation distinguishes composite
coherence from ordinary phase alignment without requiring a
thermodynamic symmetry-breaking interpretation.

On a magnetized background, the marginal linear mode carries both
ordinary and spin-weighted coherence. The amplitude-ratio comparison
therefore tests a prediction about mode composition near onset.
The pointwise intervals include the linear ratio at both
sampled couplings adjacent to the analytical onset for all three sizes,
whereas the intervals at \(K=1.095\) lie above it
(Fig.~\ref{fig:magnetic}). This local quantitative consistency is
meaningful for a prediction derived at linear onset. It neither requires
agreement over the full coupling range nor makes the finite-time
amplitude ratio a direct measurement of the growing eigenvector.

The rate-dependent crossing tests address a separate prediction and
retain a different inferential status. At \(\tau=0.75\) and \(1\),
both conditional outer-minus-middle contrasts have the sign predicted
for a finite-rate dip, but the prespecified crossing-support requirement
fails. At \(\tau=1.5\), the conditional intervals include zero and the
outer point crossings are ambiguous. All three direct tests remain
unresolved under the complete decision rule (Table~\ref{tab:verdict}).
This outcome establishes neither a numerical minimum nor a numerical
refutation of the analytical criterion. The accepted local-fit median
rates lie near the evaluated theoretical minima, but their low acceptance
fractions and conditioning prevent an independent detection or an
unconditional measurement of the optimum (Numerical Supplement,
Sec.~S5). Their fitted levels are absolute pseudocritical couplings,
not optimized gains relative to a baseline.

The scope of these comparisons is set by the numerical design.
The \(N^{-1/4}\) scaling is a working assumption, the crossing topology
depends on the finite-size curves, and no controlled \(N\to\infty\)
extrapolation is available. The threshold-level time-step screens remain
unresolved, and convergence with burn duration, sampling duration and
initialization has not been established. These limitations restrict the
interpretation of the finite-size crossings without removing the
direct mode-structure observations or the local magnetic comparison.
The analytical result itself concerns a mean-field linear instability
threshold: it gives sufficient, not necessary, conditions and does not
establish optimizer uniqueness or a nonlinear phase boundary.

\section{Conclusion}\label{sec:conclusion}

We derived the composite-mode linear instability threshold of a dynamic
Ising--Kuramoto model in terms of the free phase correlation and the
spin-switching response. Under the stated assumptions, two strict
inequalities ensure that this threshold attains a global minimum at a
finite positive switching rate, strictly below both endpoint thresholds.
The low-rate condition depends on integrated phase memory, while the
high-rate condition depends on the short-time correlation structure.
Finite-\(\tau\) OU forcing satisfies the latter condition for
\(\mu>0\); applying the sufficient criterion still requires checking
the former. The white-noise threshold is monotone or exactly flat.

The sampled finite systems exhibit composite coherence with an antiphase
sector bias, and the magnetized amplitude ratios provide a local
quantitative check near the predicted linear onset. Direct finite-size
tests of a nonmagnetic threshold minimum remain unresolved under the
prespecified support rule, despite the predicted conditional contrast
signs at two OU times. The principal analytical result is the sufficient
correlation-function criterion that separates the integrated and
short-time sufficient conditions for a finite positive global threshold minimum.

\section*{Data and code availability}

The simulation and analysis code, realization-level summaries, processed
data, bootstrap outputs, figure inputs, and reproducibility records
supporting this study are archived on Zenodo at
\url{https://doi.org/10.5281/zenodo.24163856}.
The minimal release excludes collective-observable time-series arrays
and particle-level histories.

\appendix
\section{Nonmagnetic first-harmonic response}\label{app:nonmag}

Treat \(A(t)=\delta Z_m(t)\) as an imposed field and impose
self-consistency after taking the response. The free process has a
uniform phase offset independent of the phase forcing and spin path.
For \(u\geq0\) put \(b=A(t-u)e^{-i\theta_0(t-u)}\). In either angular
projection the term proportional to
\(A(t-u)^* e^{i[\theta_0(t)+\theta_0(t-u)]}\) averages to zero under a
common rotation. Consequently
\begin{align}
 i\av{e^{i\theta_0(t)}\operatorname{Im} b}
 &=\av{e^{i\theta_0(t)}\operatorname{Re} b}\nonumber\\
 &=\tfrac12 A(t-u)G(u). \label{eq:project}
\end{align}
For an imposed field \(h(t)\), the two-state master equation following
from Eq.~\eqref{eq:flip} gives the mean spin \(q\),
\begin{equation}
 \dot q=-\gamma q+\gamma\tanh(\beta_s h). \label{eq:q}
\end{equation}
In particular, at zero field the conditional mean after a lag \(u\)
is \(e^{-\gamma u}\) times the initial spin. Thus
\(\av{\sigma(t)\sigma(t-u)}=e^{-\gamma u}\).

For the direct phase response, integrate the first-order drift
\(\delta\dot\theta=\mu K\sigma\operatorname{Im}(Ae^{-i\theta_0})\)
along a free trajectory. In \(Z_m\), the spin at the observation time
pairs with the spin in this drift. Their correlation and
Eq.~\eqref{eq:project} give the retarded response
\begin{equation}
 \delta Z_m^{(\theta)}(t)=\frac{\mu K}{2}
       \int_0^\infty e^{-\gamma u}G(u)A(t-u)\,\dd u . \label{eq:thetares}
\end{equation}
For the spin response, hold the free phase path fixed. Its imposed
first-harmonic field is \(\delta h=K\operatorname{Re}(Ae^{-i\theta_0})\).
The angular average of this perturbation is zero, so \(\delta m=0\).
Solving Eq.~\eqref{eq:q} to first order gives
\(\delta q(t)=\beta_s\gamma\int_0^\infty e^{-\gamma u}\delta h(t-u)\,\dd u\).
Averaging \(e^{i\theta_0(t)}\delta q(t)\) using
Eq.~\eqref{eq:project} yields
\begin{equation}
 \delta Z_m^{(\sigma)}(t)=\frac{\beta_s\gamma K}{2}
        \int_0^\infty e^{-\gamma u}G(u)A(t-u)\,\dd u . \label{eq:spinres}
\end{equation}
Setting \(A=\delta Z_m^{(\theta)}+\delta Z_m^{(\sigma)}\) and taking
\(A(t)\propto e^{st}\) recovers Eq.~\eqref{eq:char}.

For \(\operatorname{Re}s\geq0\), positivity gives
\(\left|\chi_{\gamma+s}\right|\leq\chi_{\gamma+\operatorname{Re}s}
\leq\chi_\gamma\). At nonzero imaginary frequency the first bound is
strict since \(G\) is positive on an interval. Thus no root in the
closed right half-plane occurs below Eq.~\eqref{eq:kc}, and \(s=0\)
is the first marginal root. Above threshold the response along the
real \(s\)-axis decreases continuously from a value greater than one
to zero and hence admits a growing mode.

For OU forcing the substitution \(x=e^{-t/\tau}\) in
Eq.~\eqref{eq:chi} gives the incomplete-gamma representation also
encountered for drifting oscillator frequencies~\cite{rougemont2006}.
With \(\alpha=\tau(D+a)\) and \(\operatorname{Re}\alpha>0\),
\begin{align}
 \chi_a&=\frac{\tau e^{D\tau}}{2}
             \int_0^1 x^{\alpha-1}e^{-D\tau x}\,\dd x\nonumber\\
 &=\frac{\tau e^{D\tau}}{2(D\tau)^\alpha}
             \gamma_{\rm inc}(\alpha,D\tau), \label{eq:ouexact}
\end{align}
where \(\gamma_{\rm inc}\) denotes the lower incomplete gamma function.

\section{Proof of the finite-rate minimum criterion}\label{app:minimum}

Since \(G(t)\) and \(tG(t)\) are integrable, they dominate the
integrand for \(M_0\) and its derivative in absolute value down to
\(\gamma=0\). Dominated convergence gives \(M_0'=-M_1\), with a right
derivative at the endpoint. Hence
\begin{align}
 F'(\gamma)&=\beta_sM_0(\gamma)-(\mu+\beta_s\gamma)M_1(\gamma),\nonumber\\
 F'(0^+)&=\beta_sM_0(0)-\mu M_1(0). \label{eq:Fprime}
\end{align}
In particular \(F\) is continuous on \([0,\infty)\). Dividing its
stationarity equation by \((\mu+\beta_s\gamma)M_0(\gamma)>0\)
gives Eq.~\eqref{eq:balance}.

For the large-rate derivative we expand the two integrals in
Eq.~\eqref{eq:Fprime} separately. Choose \(\delta>0\) such that
\(|G(t)-1-gt|\leq C t^2\) on \([0,\delta]\). For \(n=0,1\) the
integrated remainder there is bounded by
\(C(n+2)!\,\gamma^{-n-3}\). The contribution beyond \(\delta\) to
\(M_n\) is at most \(e^{-\gamma\delta}M_n(0)\). Extending the
polynomial terms to infinity also costs an exponentially small tail.
Termwise integration gives
\begin{align}
 M_0(\gamma)&=\gamma^{-1}+g\gamma^{-2}+O(\gamma^{-3}),\nonumber\\
 M_1(\gamma)&=\gamma^{-2}+2g\gamma^{-3}+O(\gamma^{-4}). \label{eq:L}
\end{align}
Substitution in Eqs.~\eqref{eq:F} and \eqref{eq:Fprime} gives, with
\(c=\mu+\beta_s g\),
\begin{align}
 F(\gamma)&=\beta_s+c\,\gamma^{-1}+O(\gamma^{-2}),\nonumber\\
 F'(\gamma)&=-c\,\gamma^{-2}+O(\gamma^{-3}). \label{eq:tail}
\end{align}
 
Under Eq.~\eqref{eq:margins}, choose \(R>0\) so that \(F'<0\) on
\([R,\infty)\). Continuity gives a maximum of \(F\) on \([0,R]\).
The positive right derivative at zero and the negative derivative at
\(R\) exclude both endpoints. Beyond \(R\), \(F\) decreases strictly
to \(\beta_s\). Thus some \(\gamma_\ast\in(0,R)\) maximizes \(F\)
globally, with \(F(\gamma_\ast)>\max\{F(0),\beta_s\}\).
Taking reciprocals yields a global minimum of \(\kc\) below both
\(2/[\mu M_0(0)]\) and \(2/\beta_s\). It is also an interior local
minimum and obeys \(F'(\gamma_\ast)=0\). The endpoint signs place
no uniqueness condition on the maximizer.

If either inequality in Eq.~\eqref{eq:margins} is an equality, its
leading derivative test is inconclusive. At low rate one needs
further information about the moments to decide the initial trend.
At \(c=0\) the sign at high rate requires information beyond the
linear term in the short-time expansion. Equation~\eqref{eq:white}
resolves the white-noise equality case exactly.

\section{Response on the magnetized branch}\label{app:magnet}

At \(Z_m=0\) the uniform magnetization follows Eq.~\eqref{eq:q} with
\(h=Jm\). Its linear relaxation rate about \(m_0\) is
\(\gamma[1-\beta_s J(1-m_0^2)]\), which gives the stability condition
in Eq.~\eqref{eq:m0}. In the first angular harmonic \(\delta m=0\).
At the fixed background field \(Jm_0\), the conditional mean spin is
\(m_0+[\sigma(t-u)-m_0]e^{-\gamma u}\). Thus the full two-time spin
correlation is
\begin{equation}
 C_\sigma(u)=m_0^2+(1-m_0^2)e^{-\gamma u}. \label{eq:Cspin}
\end{equation}
For \(A(t)\propto e^{st}\), replacing the spin correlator in
Eq.~\eqref{eq:thetares} by \(C_\sigma(u)\) yields
\begin{equation}
 \frac{\delta Z_m^{(\theta)}}{A}
      =\mu K[m_0^2\chi_s+(1-m_0^2)\chi_{\gamma+s}]. \label{eq:magphase}
\end{equation}
The field derivative of \(\tanh(\beta_s h)\) at \(h=Jm_0\) is
\(\beta_s(1-m_0^2)\). Equation~\eqref{eq:spinres} therefore gives
\(\delta Z_m^{(\sigma)}/A=\beta_s\gamma K(1-m_0^2)\chi_{\gamma+s}\).
Adding the responses gives the characteristic equation
\begin{equation}
 1=K\big[\mu m_0^2\chi_s+
 (1-m_0^2)(\mu+\beta_s\gamma)\chi_{\gamma+s}\big]. \label{eq:magchar}
\end{equation}
Both susceptibilities enter with positive weights. The positivity
argument in Appendix~\ref{app:nonmag} again gives stationary onset,
and \(s=0\) yields Eq.~\eqref{eq:magkc}.

In \(Z\) the observed phase carries unit weight. Only the direct phase
response contributes, with the mean spin \(m_0\) in the drift.
Equation~\eqref{eq:project} then gives
\(\delta Z=\mu K m_0\chi_s A\). At stationary onset the signed ratio
of eigenvector components is
\(\delta Z/\delta Z_m=\mu\kc^{(m_0)}m_0\chi_0\). Taking its
magnitude yields Eq.~\eqref{eq:magratio} on either branch.

\bibliography{references}
\end{document}


\begin{center}
{\large\bfseries Numerical Supplement}\\[4pt]
Memory-induced optimal switching in a dynamic Ising--Kuramoto model
\end{center}
\section{Nonmagnetic campaign and sampling design}\label{supp:designA}

The nonmagnetic campaign used \(J=0.4\), \(\beta_s=2\), \(D=\mu=1\), and
\(\lambda=0\). The initial phases were independent and uniform on the
circle, spins were independent and unbiased, and each initial OU force
was drawn from \(\mathcal N(0,D/\tau)\). Each step sampled the joint
conditional law of the OU endpoint and its time integral. Both stages
of the Heun phase update used that same integral and the old spins.
We wrapped the corrected phases modulo \(2\pi\), evaluated the flip rates
there with the old spins and self-excluded fields, and applied simultaneous
independent Bernoulli flips with probabilities \(1-\exp(-w_i\Delta t)\).
This splitting approximates the coupled dynamics at finite \(\Delta t\).

\begin{center}
\begin{minipage}{\linewidth}
\centering\small
\refstepcounter{table}\label{stab:a_design}
\textbf{Table \thetable.} Trajectory inventory for \(J=0.4\).
All stages used burn time 200, sample time 500 and recording interval 0.05.
Each trajectory contributed one time mean to the analysis of a given
observable. Replicate counts apply per sampled parameter point.\par\smallskip
\begin{tabular}{@{}lccrr@{}}
\toprule
Stage & \(N\) & \(\Delta t\) & Replicates & Trajectories\\
\midrule
Threshold step screen & 1000, 2000 & 0.005, 0.0025 & \(6\to12\) & 1872\\
Composite-mode observations & 500, 1000, 2000 & 0.0025 & 12 & 252\\
Primary crossing comparison & 1000, 2000 & 0.0025 & 12 & 6264\\
Size-pair check & 500 & 0.0025 & 12 & 1332\\
\midrule
Total & & & & 9720\\
\bottomrule
\end{tabular}
\end{minipage}
\end{center}

For the composite-mode panel we fixed \(\tau=\gamma=1\) and sampled
\(K=0.920,0.925,\ldots,0.950\) at each of the three sizes.
Table~\ref{stab:a_roots} lists the separate primary design.
At each \(\tau\), the first, fourth and seventh rates defined the
low, middle and high points used for the direct shape test and the
\(N=500\) size check. The five consecutive interior rates defined the
window for the conditional fits. All three selections preceded production.

For the crossing grids let \(K_j=K_0+0.005j\), where \(K_0\) is the analytical
threshold at the specified \((\tau,\gamma)\), rounded to 12 decimal places
with ties to even. Each initial grid had \(j=-3,\ldots,3\). A root or
plateau within one grid step of an edge requested an extension on that
side. A missing or ambiguous crossing, no exchange of ordering, or a
valid bootstrap fraction below 0.90 requested extensions on both sides.
Each round added at most one step on each requested side, through a maximum
of \(|j|=6\), retaining the same replicates. Requests from the sizes or
step lengths being compared were combined. The \(N=500\) checks started
on the primary grids. Their prescribed extensions were mirrored at both
primary sizes. Table~\ref{stab:a_roots} gives the final inclusive index ranges.

The trajectory master seed was 20261027. We hashed the compact ASCII JSON
array containing this integer, the namespace \texttt{trajectory}, the stage
domain, \(\tau,\gamma,N\), replicate label and \(\Delta t\), in that order.
After the master seed, entries were strings. Real parameters used canonical
decimal strings without trailing zeros. The SHA-256 digest, read as an
unsigned little-endian integer, seeded NumPy's PCG64 generator. Stage
domains separated the step screen, composite-mode panel and primary
production. The \(N=500\) size check used the production domain.
Excluding \(K\) from the key paired all couplings on a replicate curve.
Distinct sizes, rates, correlation times, step lengths and domains used
separate streams. Production used labels 0 through 11. The screen
retained its initial labels 0 through 5 when extended to 12.

For the uncertainty display in main Figure~2, we resampled the 12
realization curves 5000 times within each size, preserving all seven
couplings and both observables jointly in each draw. The PCG64 streams
used NumPy's \texttt{SeedSequence([20261106, N, 2])}, separately for
each size. The pointwise intervals are the 2.5th and 97.5th percentiles
with linear quantile interpolation. The plotted centers retain the
original processed means. Time samples were not treated as independent
observations. These intervals describe finite-seed variability at the
recorded settings, not simultaneous bands or systematic errors; their
overlap is not a formal test of size-ordering differences. This
uncertainty display does not alter the crossing or fit bootstraps below.

\section{Crossing support and fixed-rate contrasts}\label{supp:supportA}

For size \(N\) and coupling \(K\), the plotted diagnostic uses the
equally weighted mean over trajectories of the time-averaged magnitude
\(r_m\). We multiply by \(N^p\), with \(p=1/4\) as a working finite-size
assumption. At each size the primary algorithm fits a nondecreasing
sequence across the sampled \(K\)'s by equal-weight isotonic regression
(the pool-adjacent-violators algorithm, PAVA). It linearly interpolates
the difference of the two resulting curves. Differences of modulus at most
\(10^{-12}\) count as zero. Connected zero components separated by at most
\(10^{-12}\) in \(K\) are merged. One isolated root or one contiguous
plateau gives a valid estimate, with the midpoint used for a plateau.
No component gives a missing estimate, and multiple disjoint components
give an ambiguous estimate. Only valid point estimates appear in the
main threshold comparison.

For each size and condition we resampled 12 replicate curves with replacement
5000 times, retaining the joint \(K\) coordinates in each draw. Bootstrap
streams used master seed 20261028, the namespace
\texttt{bootstrap-crossing}, the condition key and the size. Root finding was
repeated per draw. The interval uses the 2.5th and 97.5th percentiles
(linear quantile interpolation) of the valid roots. Support requires a
valid point root, an exchange of ordering, a root component more than one
grid step from each edge, and at least 0.90 valid bootstrap fraction.
These criteria also apply to conditions whose grid has reached its cap.

{\small\setlength{\tabcolsep}{4pt}
\begin{longtable}{@{}rrcrcrrr@{}}
\caption{Final primary \(K\) grid and crossing support for \(N=1000,2000\).
An asterisk marks a point satisfying the full support criterion. A denotes
multiple disjoint roots in the point-mean curves. The interval is conditional
on a valid crossing and may exist even for an ambiguous point estimate.
The last three columns give valid and ambiguous draw counts and the valid
fraction. All 5000 draws per row are accounted for by these two counts.
Missing and nonfinite draws and valid plateaus each numbered zero.}
\label{stab:a_roots}\\
\toprule
\(\tau\)&\(\gamma\)&\(j\) range&\(K_\times\)&Conditional 95\% interval&
Valid&Ambig.&Fraction\\\midrule
\endfirsthead
\toprule
\(\tau\)&\(\gamma\)&\(j\) range&\(K_\times\)&Conditional 95\% interval&
Valid&Ambig.&Fraction\\\midrule
\endhead
\bottomrule\endfoot
0.75 & 1.5 & $[-6,6]$ & $0.94583$ & $[0.94134,0.95373]$ & 2828 & 2172 & 0.5656 \\
0.75 & 2.2 & $[-6,6]$ & $0.94600$ & $[0.94427,0.95012]$ & 2656 & 2344 & 0.5312 \\
0.75 & 2.45 & $[-6,6]$ & A & $[0.92929,0.94015]$ & 3172 & 1828 & 0.6344 \\
0.75 & 2.629 & $[-6,6]$ & $0.93358^{*}$ & $[0.93036,0.93971]$ & 4858 & 142 & 0.9716 \\
0.75 & 2.8 & $[-6,6]$ & A & $[0.92590,0.94489]$ & 2459 & 2541 & 0.4918 \\
0.75 & 3.1 & $[-6,6]$ & A & $[0.93157,0.94549]$ & 2860 & 2140 & 0.5720 \\
0.75 & 5 & $[-3,3]$ & $0.95085^{*}$ & $[0.94251,0.95299]$ & 4572 & 428 & 0.9144 \\
\midrule
1 & 0.5 & $[-6,6]$ & $0.97031^{*}$ & $[0.96428,0.97814]$ & 4574 & 426 & 0.9148 \\
1 & 1.4 & $[-6,6]$ & $0.92085$ & $[0.91201,0.92902]$ & 3817 & 1183 & 0.7634 \\
1 & 1.65 & $[-6,6]$ & A & $[0.90815,0.92467]$ & 2420 & 2580 & 0.4840 \\
1 & 1.826 & $[-6,6]$ & $0.90725$ & $[0.90321,0.91746]$ & 3562 & 1438 & 0.7124 \\
1 & 2 & $[-3,3]$ & $0.92146^{*}$ & $[0.91458,0.92762]$ & 4618 & 382 & 0.9236 \\
1 & 2.3 & $[-6,6]$ & $0.91358$ & $[0.90479,0.91951]$ & 2958 & 2042 & 0.5916 \\
1 & 5 & $[-6,6]$ & $0.93806^{*}$ & $[0.93402,0.94393]$ & 4855 & 145 & 0.9710 \\
\midrule
1.5 & 0.5 & $[-6,6]$ & A & $[0.88679,0.91039]$ & 1644 & 3356 & 0.3288 \\
1.5 & 0.75 & $[-6,6]$ & $0.88354$ & $[0.87027,0.89381]$ & 3072 & 1928 & 0.6144 \\
1.5 & 0.9 & $[-6,6]$ & $0.87243$ & $[0.86711,0.88122]$ & 2782 & 2218 & 0.5564 \\
1.5 & 1.024 & $[-6,6]$ & $0.87959$ & $[0.87389,0.89090]$ & 4309 & 691 & 0.8618 \\
1.5 & 1.15 & $[-6,6]$ & $0.87448$ & $[0.86496,0.88306]$ & 2332 & 2668 & 0.4664 \\
1.5 & 1.35 & $[-6,6]$ & $0.87220$ & $[0.85893,0.88076]$ & 3303 & 1697 & 0.6606 \\
1.5 & 2 & $[-6,6]$ & A & $[0.88389,0.89598]$ & 2515 & 2485 & 0.5030 \\
\end{longtable}}

Fifteen point estimates are valid and six are ambiguous. Five satisfy
the full support requirement. For the direct contrasts, bootstrap
draws at different rates were independent. Each side-minus-middle
difference used only draw indices with valid crossings at both rates.
The retained counts for low-minus-middle and high-minus-middle were
\((2757,4439)\), \((3255,3453)\), and \((1403,2178)\) out of 5000
for \(\tau=0.75,1,1.5\), respectively. Both lower interval endpoints
had to exceed zero and all three individual crossings had to satisfy
the full support rule. All three tests were unresolved.

\section{Sensitivity to the working exponent, interpolation and size pair}
\label{supp:sensitivityA}

Table~\ref{stab:a_est} gives the prespecified point-estimate comparisons.
For example, at \(\tau=0.75,\gamma=2.45\), the primary \(p=0.25\)
curves have several roots, whereas \(p=0.20\) and 0.30 each give a
single root at different couplings. At \(\tau=1,\gamma=1.826\),
lowering \(p\) to 0.20 makes the point estimate ambiguous. At fixed
\(p=0.25\), the two raw interpolants have the same root classification
here and can yield different locations within a bracket. The primary
analysis retains \(p=0.25\), PAVA and linear interpolation throughout.

{\small\setlength{\tabcolsep}{6pt}
\begin{longtable}{@{}rrccccc@{}}
\caption{Point crossings for the primary pair \(N=1000,2000\). PAVA
columns use piecewise-linear interpolation. Raw columns at \(p=0.25\)
interpolate the unsmoothed difference by straight segments or by a
shape-preserving piecewise cubic Hermite interpolating polynomial
(PCHIP). A denotes multiple disjoint roots. Each displayed location is
a finite-grid diagnostic based on the seed means.}
\label{stab:a_est}\\
\toprule
& & \multicolumn{3}{c}{PAVA by exponent \(p\)} & \multicolumn{2}{c}{Raw, \(p=0.25\)}\\
\cmidrule(lr){3-5}\cmidrule(lr){6-7}
\(\tau\)&\(\gamma\)&0.20&0.25&0.30&Linear&PCHIP\\\midrule
\endfirsthead
\toprule
\(\tau\)&\(\gamma\)&PAVA 0.20&PAVA 0.25&PAVA 0.30&Raw linear&Raw PCHIP\\\midrule
\endhead
\bottomrule\endfoot
0.75 & 1.5 & 0.95434 & 0.94583 & A & 0.94583 & 0.94563 \\
0.75 & 2.2 & 0.94919 & 0.94600 & A & 0.94600 & 0.94622 \\
0.75 & 2.45 & 0.94250 & A & 0.92843 & A & A \\
0.75 & 2.629 & 0.94178 & 0.93358 & 0.93000 & 0.93358 & 0.93329 \\
0.75 & 2.8 & 0.94436 & A & 0.92607 & A & A \\
0.75 & 3.1 & 0.94652 & A & 0.93102 & A & A \\
0.75 & 5 & 0.95426 & 0.95085 & 0.94026 & 0.95085 & 0.95114 \\
\midrule
1 & 0.5 & 0.97434 & 0.97031 & 0.96645 & 0.97031 & 0.97033 \\
1 & 1.4 & 0.92880 & 0.92085 & 0.91247 & 0.92085 & 0.92096 \\
1 & 1.65 & 0.92651 & A & 0.90774 & A & A \\
1 & 1.826 & A & 0.90725 & 0.90357 & 0.90725 & 0.90698 \\
1 & 2 & 0.92741 & 0.92146 & 0.91477 & 0.92146 & 0.92146 \\
1 & 2.3 & A & 0.91358 & 0.90502 & 0.91358 & 0.91401 \\
1 & 5 & 0.94297 & 0.93806 & 0.93401 & 0.93806 & 0.93806 \\
\midrule
1.5 & 0.5 & A & A & 0.88856 & A & A \\
1.5 & 0.75 & 0.88776 & 0.88354 & A & 0.88354 & 0.88355 \\
1.5 & 0.9 & 0.87428 & 0.87243 & A & 0.87243 & 0.87261 \\
1.5 & 1.024 & A & 0.87959 & 0.87699 & 0.87959 & 0.87950 \\
1.5 & 1.15 & 0.88206 & 0.87448 & A & 0.87448 & 0.87413 \\
1.5 & 1.35 & 0.87998 & 0.87220 & 0.86101 & 0.87220 & 0.87204 \\
1.5 & 2 & 0.89599 & A & 0.88538 & A & A \\
\end{longtable}}

Table~\ref{stab:a_size} compares adjacent size pairs on the same grids.
The smaller pair shifts the middle-rate point estimate upward by
\(0.01268\) at \(\tau=1\) and downward by \(0.00858\) at \(\tau=1.5\).
Changing the pair also changes whether a single root exists at some
conditions. The two pairs share the \(N=1000\) data. Their differences
describe sensitivity to the sampled sizes and do not determine a
thermodynamic extrapolation.

{\small\setlength{\tabcolsep}{7pt}\setlength{\LTcapwidth}{\linewidth}
\begin{longtable}{@{}rrccrcc@{}}
\caption{Size-pair sensitivity at the nine predetermined contrast
points. Pair 1 is \(N=1000,2000\) and pair 2 is \(N=500,1000\).
\(f_i\) is the fraction of 5000 draws returning a valid crossing for
pair \(i\). A denotes an ambiguous point estimate. A dash indicates an
undefined difference.}
\label{stab:a_size}\\
\toprule
\(\tau\)&\(\gamma\)&\(K_{\times,1}\)&\(K_{\times,2}\)&
\(K_{\times,2}-K_{\times,1}\)&\(f_1\)&\(f_2\)\\\midrule
\endfirsthead
\toprule
\(\tau\)&\(\gamma\)&\(K_{\times,1}\)&\(K_{\times,2}\)&
\(K_{\times,2}-K_{\times,1}\)&\(f_1\)&\(f_2\)\\\midrule
\endhead
\bottomrule\endfoot
0.75 & 1.5 & 0.94583 & 0.95409 & $+0.00826$ & 0.5656 & 0.6438 \\
0.75 & 2.629 & 0.93358 & 0.93441 & $+0.00083$ & 0.9716 & 0.7676 \\
0.75 & 5 & 0.95085 & 0.94544 & $-0.00541$ & 0.9144 & 0.9532 \\
\midrule
1 & 0.5 & 0.97031 & A & --- & 0.9148 & 0.2410 \\
1 & 1.826 & 0.90725 & 0.91993 & $+0.01268$ & 0.7124 & 0.6478 \\
1 & 5 & 0.93806 & 0.94348 & $+0.00542$ & 0.9710 & 0.6122 \\
\midrule
1.5 & 0.5 & A & 0.89771 & --- & 0.3288 & 0.3740 \\
1.5 & 1.024 & 0.87959 & 0.87101 & $-0.00858$ & 0.8618 & 0.7672 \\
1.5 & 2 & A & A & --- & 0.5030 & 0.3800 \\
\end{longtable}}

\section{Time-step and sampling diagnostics at \(J=0.4\)}
\label{sec:sup-time}

The time-step screen used the primary size pair and the middle rate of each
direct contrast. It compared \(\Delta t=0.005\) with \(0.0025\), using the
same physical burn and observation times at both steps. Streams at the two
steps were independent. The first stage used six realizations per condition.
The protocol called for six more if the point difference exceeded \(0.010\)
in magnitude, only one step gave a unique crossing, the pointwise interval
for the difference excluded zero, or the prescribed grid extensions left
inadequate crossing support. The second stage retained the original six
realizations. A persistent trigger selected the finer step, including when
the support test remained inconclusive at both steps.

\begin{table}[htbp]
\centering
\caption{Threshold-level time-step screen at \(J=0.4\), after extension to
12 realizations. The difference is the crossing at \(0.005\) minus the
crossing at \(0.0025\). Intervals retain draws with a usable crossing at
both steps. A dash denotes an unavailable point difference.}
\label{tab:sup-dt}
\begin{tabular}{@{}rrrrl@{}}
\toprule
\(\tau\)&\(\gamma\)&Point difference&Conditional 95\% interval&Decision\\
\midrule
0.75&2.629& 0.00376&\([-0.00606,\, 0.01066]\)&Unresolved\\
1.00&1.826& 0.00403&\([-0.01083,\, 0.01012]\)&Unresolved\\
1.50&1.024& ---    &\([-0.01318,\, 0.01081]\)&Unresolved\\
\bottomrule
\end{tabular}
\end{table}

All three comparisons remained unresolved at the grid cap
(Table~\ref{tab:sup-dt}). In the third, only one step had a usable point
crossing. The selection rule gave \(\Delta t=0.0025\) for production.
These intervals quantify the seed variability of the crossing difference
under the specified screen. They leave the continuum-limit bias undetermined.

Each run retained means for the first and second halves of the sampling
period, 20 equal-duration block means and, when defined, an integrated
autocorrelation time and effective sample count. Circular observables were
diagnosed on their sine and cosine components. Records with missing values
or zero variance have an undefined autocorrelation-time estimate, preserving
the original time grid. For the 7596 production crossing runs,
the smallest and median effective counts for \(r_m\) were 8.73 and 57.61
among 10\,000 records per run. The largest absolute difference between
half-run means of \(r_m\) was 0.1302. These summaries describe temporal
variability over the sampled duration. Burn-time dependence, observation-time
dependence and sensitivity to initialization remain undetermined. All runs
entered the prespecified analysis irrespective of these diagnostics.

\section{Conditional fits of the pseudocritical curves}
\label{sec:sup-fits}

For each bootstrap draw, an unweighted quadratic was fit to the usable
crossings in the five-rate window fixed before production. The fit used
ordinary least squares in an affinely rescaled rate. Acceptance required at
least three finite crossings, finite fitted quantities, positive curvature
and a vertex in the original five-rate interval. The same rule was applied
to the crossing estimates obtained directly from the seed means.
Table~\ref{tab:sup-fitbranches} includes every draw, while
Table~\ref{tab:sup-fitvalues} summarizes the accepted draws.

\begin{table}[htbp]
\centering
\caption{Fate of all 5000 draws per OU time in the local fits. The four
middle columns give mutually exclusive failure counts, with \(a\) the
quadratic coefficient. The point-fit
column applies the same rule to the original seed-mean crossing curve.}
\label{tab:sup-fitbranches}
\small
\begin{tabular}{@{}rlrrrrrr@{}}
\toprule
\(\tau\)&Point fit&Too few&Nonfinite&\(a\le0\)&Outside&Accepted&Fraction\\
\midrule
0.75&Too few crossings&1223&0&589&196&2992&0.5984\\
1.00&Accepted&754&0&1485&375&2386&0.4772\\
1.50&Vertex outside&1288&0&1735&416&1561&0.3122\\
\bottomrule
\end{tabular}
\end{table}

\begin{table}[htbp]
\centering
\caption{Local quadratic vertex and absolute pseudocritical level,
conditional on fit acceptance. Each entry gives the median and the
pointwise 2.5th and 97.5th percentiles of the corresponding accepted
draws. The fixed-rate direct test remains unresolved in every row.}
\label{tab:sup-fitvalues}
\begin{tabular}{@{}rrlrl@{}}
\toprule
\(\tau\)&\multicolumn{2}{c}{Vertex rate}&
\multicolumn{2}{c}{Absolute fitted \(K_{\min}\)}\\
\midrule
0.75&2.71474&[2.46929, 2.97985]&0.93201&[0.92752, 0.93721]\\
1.00&1.79359&[1.61594, 2.21151]&0.91028&[0.89943, 0.91763]\\
1.50&1.10667&[0.87252, 1.31890]&0.87321&[0.86364, 0.87922]\\
\bottomrule
\end{tabular}
\end{table}

Conditional on acceptance, the median vertex rates
\(2.71474,1.79359,1.10667\) lie near the evaluated analytical minima
\(2.629,1.826,1.024\), respectively. The acceptance fractions are
\(0.5984,0.4772,0.3122\), and all three direct tests remain unresolved;
this descriptive comparison is not an unconditional measurement of
\(\gamma_\ast\). The fitted ordinate is an absolute pseudocritical level,
not a measured optimized gain.

\begin{figure}[htbp]
\centering
\includegraphics[width=0.93\linewidth]{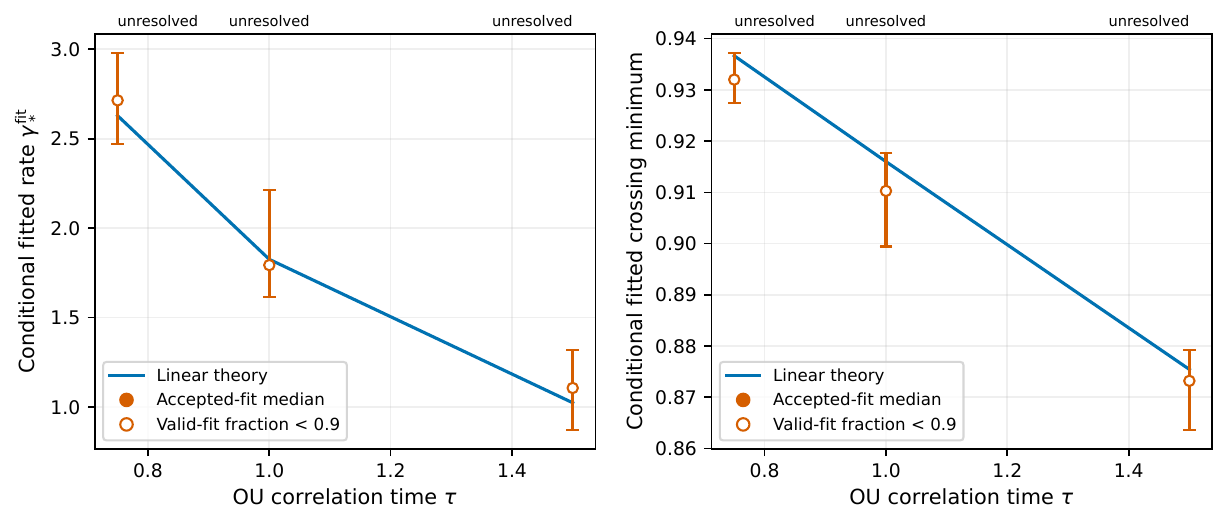}
\caption{Local fits of the finite-size crossings at \(J=0.4\),
\(\beta_s=2,\ D=\mu=1,\ \lambda=0\).
The blue segments connect the three evaluated minima of the analytical
linear threshold as a guide to the eye. Left, the switching rate at
the vertex. Right, the absolute threshold level there. Orange circles
are simulation-based medians from accepted quadratic fits in the fixed
five-rate windows, using \(N=1000,2000\) crossings with the working
\(N^{-1/4}\) assumption. Bars span the pointwise 95\% intervals conditional
on acceptance. The accepted-fit median rates lie near the evaluated
analytical minima. All circles are open because fewer than 90\% of draws
were accepted for each \(\tau\). The direct numerical tests of
nonmonotonicity remain unresolved for all three times. These conditional
summaries are not unconditional measurements of the optimizer; the
fitted levels are absolute pseudocritical couplings, not measured
optimized gains relative to a baseline.}
\label{fig:sup-fit}
\end{figure}

\clearpage
\section{Magnetized-background ratio at \(J=0.8\)}
\label{sec:sup-mag}

\begin{table}[htbp]
\centering
\caption{Production protocol for the magnetized-background amplitude ratio.
Each of the 15 size--coupling conditions contributes 20 trajectories.}
\label{tab:sup-mag}
\begin{tabular}{@{}ll@{}}
\toprule
Quantity&Recorded setting\\
\midrule
Model&\(J=0.8,\ \beta_s=2,\ D=\tau=\gamma=\mu=1,\ \lambda=0\)\\
System sizes&\(N=500,1000,2000\)\\
Couplings&\(K=1.095,1.100,1.105,1.110,1.115\)\\
Random streams&PCG64, integer seeds \(0,\ldots,19\) reused at each \(N,K\)\\
Time step&0.005\\
Burn and sample times&200 and 500 (40\,000 and 100\,000 updates)\\
Recording stride&10 updates, or 0.05 time units, for 10\,000 records\\
Initialization&Uniform phases, unbiased spins, stationary Gaussian OU noise\\
Ratio bootstrap&10\,000 replicates, master seed 246026\\
\bottomrule
\end{tabular}
\end{table}

The finite-\(N\) spin field excluded the self term. Each step sampled the OU
endpoint and time integral jointly and used that integral in both stages
of the Heun phase update. The simultaneous spin update used the corrected
phases and the old spins. For each \(N,K\), the ratio was
\(\sum_j \overline r_j/\sum_j\overline{r_{m,j}}\), where a bar denotes
a trajectory time average and \(j\) indexes the 20 seeds. The bootstrap
resampled numerator and denominator together, retaining each sampled
seed label across all five \(K\) at fixed \(N\). Intervals are nominal
pointwise 95\% percentile intervals. Reuse of labels across sizes and
couplings precludes treating all 300 trajectories as mutually independent.

A separate pilot used \(N=500,\ K=1.105\), seeds \(0,\ldots,5\), and
\(\Delta t=0.010,0.005,0.0025\) at the same physical burn and sample times.
The fixed stride of ten gave recording intervals of \(0.10,0.05,0.025\).
For each of \(\overline r,\overline r_m\), their ratio of means, and
mean \(|m|\), the pilot compared \(0.005\) with \(0.0025\). The bound
for flagging a difference was the larger of 2.5 paired bootstrap standard
errors and a prespecified floor (respectively \(0.010,0.010,0.001,0.002\)).
All four quantities passed. The ratio difference was \(-0.00003236\)
with standard error 0.002837 (10\,000 resamples, seed 261026). The
production choice was \(0.005\). Pairing was by integer label across
steps, with distinct Brownian paths. The six labels also occur in
production. This single-condition pilot leaves the systematic error
at the \(10^{-3}\) scale unresolved.

The linear reference is \(0.846193\) at \(\kc^{(m_0)}=1.1058624\).
The pointwise intervals at \(K=1.105\) and \(1.110\) enclose
the reference at all three sizes. At \(K=1.095\), all three lie above
it. The observed local consistency concerns a ratio of finite-time
mean amplitudes near onset. Across the 300 production runs the
effective counts for \(r_m\) had a minimum of about 9.7 and a median
of 30.5. The intervals describe seed variability at the recorded time
step, sizes, initialization and run length.

\clearpage
\section{Numerical records and figure provenance}
\label{sec:sup-records}

The following paths are relative to the indicated campaign directories.
They identify the recorded configurations, per-run outputs, analysis and
plot inputs for the figures in the present numbering.
These local paths identify provenance, not external download locations.
The corresponding public data-and-code release is available on Zenodo at
\url{https://doi.org/10.5281/zenodo.24163856}.

\begin{description}
\item[\(J=0.4\).] In \nolinkurl{clean_campaign_20261027}, the frozen
input manifest is \nolinkurl{protocol/FREEZE.json}. The configuration,
selection plan and seed mapping are in \nolinkurl{protocol/campaign_config.json},
\nolinkurl{protocol/analysis_plan.md} and \nolinkurl{protocol/seed_scheme.md}.
The tested environment is in \nolinkurl{environment/environment.json}.
The runner and simulator are \nolinkurl{source/run_campaign.py} and
\nolinkurl{source/dynamic_spin_ou.py}. Per-run records are in
\nolinkurl{raw/tasks/}, with a combined table at
\nolinkurl{seed_summaries/all_runs.csv}. The numerical estimators and
reporting code are in \nolinkurl{source/analysis_core.py} and
\nolinkurl{source/postprocess.py}. The retained draw arrays are in
\nolinkurl{bootstrap/}, and the sampling summary is in
\nolinkurl{diagnostics/stationarity_summary.csv}.
\item[Figures 1 and 2.] In the same directory, Figure~1 uses
\nolinkurl{processed/fig3_clean_crossings.csv} and
\nolinkurl{protocol/theory_curves.csv}. Figure~2 uses
\nolinkurl{processed/fig1_clean.csv}. Their original plotting code is
\nolinkurl{source/postprocess.py}, with plot inputs documented in
\nolinkurl{figures/FIGURE_INPUTS.json}. The direct-test intervals and
decisions are in \nolinkurl{processed/direct_nonmonotonicity.csv}.
\item[Figure~\ref{fig:sup-fit}.] The local fits are in
\nolinkurl{processed/fixed_window_fits.json} and
\nolinkurl{processed/fig4_conditional_fits.csv}. The theory reference is
\nolinkurl{protocol/theory_optima.csv}. The plotted version uses
\nolinkurl{publication_layout/figure4_layout_only.py}, with the numeric
display check in \nolinkurl{publication_layout/LAYOUT_CHANGE_RECORD.json}.
The trajectory checks and completion counts are recorded in
\nolinkurl{reports/FINAL_PROVENANCE_INDEPENDENT_20261104.json},
alongside \nolinkurl{reports/REPRODUCIBILITY_MANIFEST.json}.
\item[\(J=0.8\) and Figure 3.] In \nolinkurl{fig2_clean_rerun_20261026},
the configuration, source manifest and environment are
\nolinkurl{fig2_clean_config.json}, \nolinkurl{SOURCE_MANIFEST.json}
and \nolinkurl{environment.json}. The runner and simulator are
\nolinkurl{source/run_campaign.py} and \nolinkurl{source/dynamic_spin_ou.py}.
The source table is \nolinkurl{data/raw/fig2_clean_seed_summaries.csv}.
\nolinkurl{analysis/build_fig2_processed.py} produces
\nolinkurl{data/processed/fig2_clean_processed.csv}. The theory check is in
\nolinkurl{validation/theory_results.json}. These two inputs feed
\nolinkurl{figures/plot_refinement/plot_variants.py}, whose markers-only
version is \nolinkurl{figures/plot_refinement/figure2_variant_A_markers_only.pdf}.
The plot verification is recorded in
\nolinkurl{figures/plot_refinement/NUMERICAL_IDENTITY_CHECK.json}.
\item[Presentation revisions.] Relative to the manuscript source directory,
\nolinkurl{minor_revision_20261106/figure2/} contains
\nolinkurl{build_figure2_uncertainty.py},
\nolinkurl{figure2_seed_intervals.csv}, the selected realization records,
the retained bootstrap draws and \nolinkurl{FIGURE2_UNCERTAINTY_CHECKS.json}.
These add the main Figure~2 intervals without changing its means.
\nolinkurl{minor_revision_20261106/legends/} contains
\nolinkurl{update_legends.py} and \nolinkurl{LEGEND_ONLY_CHECKS.json}
for the legend-only revisions to main Figure~1 and
Figure~\ref{fig:sup-fit}; their numerical content is unchanged.
\end{description}